\documentclass[12pt, letterpaper, oneside]{article}

\usepackage{graphicx}
\usepackage{epstopdf}
\usepackage[body={6.5in, 9in}, right=1in, top=1in]{geometry}
\usepackage{amssymb} 
\usepackage{amsmath} 
\usepackage{hyperref}
\usepackage{sidecap}

\newcommand\ee{\end{equation}}
\newcommand\be{\begin{equation}}
\newcommand\eea{\end{eqnarray}}
\newcommand\bea{\begin{eqnarray}}
\newcommand{\sfrac}[2]{{\textstyle\frac{#1}{#2}}}
\newcommand\di{\partial}

\begin{document}

\begin{center}
\Large{\textbf{Two-dimensional Lorentz invariance \\[.1cm]
of spherically symmetric black holes }} \\[.8cm]
\large{Lam Hui and Alberto Nicolis}
\\[0.4cm]
\small{\textit{ Department of Physics and ISCAP, \\ 
Columbia University, New York, NY 10027, USA}}

\end{center}

\vspace{.2cm}

\begin{abstract}
We show that a static spherically symmetric black hole,
in a generic theory of gravity with generic matter fields,
has a two-dimensional Lorentz symmetry.
\end{abstract}

\section{Introduction}
As is well known, the Schwarzschild solution of general relativity (GR) admits a regular extension ``past the horizon"---the so-called Kruskal extension. Its regularity is made manifest by using Kruskal coordinates $X,T$ in place of the Schwarzschild radial and time coordinates, so that the metric reads
\footnote{We are using a dimensionful version of Kruskal coordinates, with units of length, so that for instance with respect to Wald's choice \cite{Wald} we have
\be
(X,T)_{\rm here} = r_S \cdot (X,T)_{\rm Wald} \; .
\ee}
\be \label{Kruskal classic}
ds^2 = \frac{4 r_S}{r}e^{-r/r_S}\big( -dT^2 + dX^2 \big) + r^2 d \Omega^2 \; ,
\ee
where $r_S = 2GM$ is the black hole's Schwarzschild radius, and $r$ is Schwarzschild's radial coordinate, related to Kruskal's $X$ and $T$ via
\be \label{r}
\big(\sfrac{r}{r_S} - 1 \big) e^{r/r_S} = \sfrac1{r_S^2} (X^2 - T^2) \; . 
\ee 
It is  manifest that the only singularity in the metric \eqref{Kruskal classic} is  the physical one at $r=0$, corresponding to the hyperbola
\be \label{hyperbola}
T^2 - X^2 = r_S^2 \; .
\ee
In this note, we want to focus on a different but equally manifest aspect of this rewriting of the Schwarzschild solution, which is usually glossed over: its {\em two-dimensional Lorentz invariance}. Considering Lorentz transformations acting on the $X,T$ coordinates in the obvious way, it is clear that $X$ and $T$ and their differentials appear in eqs.~\eqref{Kruskal classic} and \eqref{r} only via Lorentz invariant combinations, and as a consequence the metric is Lorentz invariant. Notice however that we don't have full two-dimensional {\em Poincar\'e} invariance---the metric depends explicitly (via $r$) on $(X^2 -T^2)$, and, as a consequence, the origin $(X,T)=0$ is a special point, and we only have Lorentz-invariance about that point. Notice also that  {\em any} regular two-dimensional metric can be cast---at least locally---into a manifestly conformally flat form. So, the truly nontrivial property of the metric \eqref{Kruskal classic} is not the appearance of the Minkowskian structure $(-dT^2 +dX^2)$, but rather the Lorentz-invariance of the  conformal factor in front of it, as well as of the $r^2$ factor multiplying the angular part of the metric. In other words: the 2D Lorentz-invariance that we are emphasizing is not a peculiarity arising for a judicious choice of the coordinates, but rather a true isometry, which can and should be talked about in a coordinate-independent way.

The isometries of the Schwarzschild solution are well known: they are $SO(3)$ rotations, and what in the original Schwarzschild coordinates takes the form of time translations. Since in going from Schwarzschild to Kruskal we are not redefining the angular coordinates, the $SO(3)$ isometries are still manifest in the $d \Omega^2$ part of \eqref{Kruskal classic}, and it must be that our 2D Lorentz isometry is nothing but Schwarzschild-time translational invariance. As a zeroth order check, the 2D Lorentz group only involves one boost generator, and is thus trivially abelian, like the group of time translations. Moreover, both groups are non-compact. They are thus isomorphic to each other.

Notice that the same relation between 2D boosts and time translations appear in 2D flat space when going from Minkowski to Rindler coordinates, respectively playing the roles of Kruskal and Schwarzschild coordinates in our discussion.  
The goal of this paper is {\em (i)} to show that for {\em any} static, spherically symmetric black hole  with finite surface gravity, there exists a set of Kruskal-like coordinates in which the metric is regular at the horizon and staticity takes the form of manifest 2D Lorentz invariance, and {\em (ii)} to provide an explicit construction of such coordinates. We will only assume regularity of the geometry at the horizon. In particular, by never using Einstein's equations, we will not commit ourselves to GR, nor will we make assumptions about the matter fields that might be present. Because of this, our results apply unaltered to modified gravity theories, as long as these admit a metric description, and spherical black hole solutions.

It should be emphasized that most of  our results are implicitly contained in previous work by Racz and Wald \cite{RW1, RW2}. So, in a sense our modest goal is to state and emphasize the Lorentz structure of their results more explicitly, and to provide a quicker derivation that we hope will be more immediate to grasp for the uninitiated (including ourselves).

\section{Setup, and desiderata}
Consider a static spherically symmetric black hole, in a generic theory of gravity with generic matter fields. We can define a Schwarzschild-like radial coordinate $r$ in such a way that the metric takes the form
\be \label{metric}
ds^2 = - f(r) dt^2 + \frac{dr^2 }{f(r)} + \rho^2(r) d\Omega^2 \; .
\ee
Without further information on the dynamics that produce such a black-hole solution, we treat $f(r)$ and $\rho(r)$ as generic functions. The horizon $r=r_h$ of the black hole corresponds to the outermost value of $r$ where $f$ vanishes,
\be
f(r_h) = 0 \; .
\ee
The form of the metric above is manifestly invariant under rotations and time translations. The latter are generated by the Killing vector
\be \label{Killing}
\xi^\mu = (1,0,0,0) \; . 
\ee
%
%

Dropping from now on the angular directions and focusing on the $(t,r)$ plane,
our goal is to replace $t$ and $r$ with a new pair of coordinates $(T,X)$ that are Kruskal-like in the sense that:
\begin{enumerate}
\item The time-translational Killing vector $\xi^\mu$ generates Lorentz transformations on $T$ and $X$;
\item The metric is conformally flat, with a Lorentz-invariant conformal factor;
\item The metric is manifestly regular at the horizon, so that $T$, $X$, and the metric can be straightforwardly extrapolated past the horizon;
\item The metric is manifestly regular at the origin $(T,X) =0 $, and our Lorentz transformations act ``around'' this origin;
\item The origin is a bifurcation point for the horizon, and the bifurcate horizon is just the light cone.
\item The $(T,X)$ coordinates cover all of the outside of the black hole;
\item In the ``inside'' direction, the $(T,X)$ coordinates cover at
  least a finite neighborhood of the horizon.
\item In the whole domain of validity of the coordinate system, $T$  is a timelike coordinate and $X$ is a spacelike one.
\end{enumerate}
These properties are all featured by the Kruskal extension of the Schwarzschild solution, and are of course highly interdependent, but not equivalent.
We will prove that they are shared by all static spherically symmetric
black-holes with finite surface gravity.

\section{Two-dimensional Lorentz invariance}
\label{2DLorentz}

Ignoring the angular variables, we will use the two-dimensional
notations $x^\mu = (t,r)$ and $X^\alpha = (T,X)$. We will refer to
these as the old and new coordinates respectively.
Our goal is to construct explicitly the $(T,X)$ coordinates where
the Lorentz invariance is manifest, and they will cover at
least a {\it finite} neighborhood of the horizon.
Consider item 1 in the list of desired properties in the previous section. Since infinitesimal 2D Lorentz transformations act as
\be
\delta X^\alpha = \omega \, \epsilon^\alpha {}_{\beta} \, X^\beta \; ,  \qquad  \epsilon^\alpha {}_{\beta} = \left( \begin{array}{cc}
0 & 1\\
1 & 0 
\end{array} \right) \; , 
\ee
where $\omega$ is an infinitesimal transformation parameter, to obey item 1 we need the components of the Killing vector in the new coordinate system to be
\be
\Xi^\alpha(X) = A \, \epsilon^\alpha {}_{\beta} \, X^\beta \; , 
\ee
where $A$ is for the moment an arbitrary (positive) constant.
Using the standard transformation law for a vector, we can rewrite this equation as a differential equation for our change of variables, $X^\alpha = X^\alpha(x)$:
\be
\xi^\mu \di_\mu X^\alpha(x) = A \, \epsilon^\alpha {}_{\beta} \, X^\beta (x) \; .
\ee
Given the simple form of $\xi^\mu$ in the original $(t,r)$ coordinates (eq.~\eqref{Killing}), the combination $\xi^\mu \di_\mu$ is simply $\di_t$, and so the most general solution to the above differential equation is
\footnote{We are implicitly demanding that $T$ be the timelike variable and $X$ the spacelike one.}
\begin{align}
T & = \psi(r) \sinh(A t + \varphi(r)) \label{T}\\
X & = \psi(r) \cosh(A t + \varphi(r)) \label{X}
\end{align}
where $\psi(r)$ and $\varphi(r)$ are generic functions.
This takes care of item 1: for any choice of $\psi$ and $\varphi$, these new coordinates get Lorentz-transformed into each other by the action of the time-translational isometry. From now we will choose 
\be
\varphi (r)= 0 \; ,
\ee
as different choices will not help us accomplish the other desiderata.

For item 2, which is {\em not} a trivial consequence of item 1, we find it more convenient to use the inverse metric. The most general Lorentz invariant inverse metric is
\be
G^{\alpha \beta} (X) = g_1(X\cdot X) \eta^{\alpha \beta} + g_2(X\cdot X) X^\alpha X^\beta \; , \qquad X \cdot X \equiv X^\alpha X^\beta \eta_{\alpha \beta} \; ,
\ee
so we want to make sure that the $g_2$ piece is absent. At generic values of $X^\alpha$, this is the same as requiring that there be no off-diagonal terms:
\be
G^{01} = g^{\mu\nu}(x) \, \di_\mu T(x) \, \di_\nu X(x) = 0 \; , 
\ee
which translates into the differential equation for $\psi(r)$
\be
-\frac{A^2}{f} \psi^2 + f \psi' {}^2 = 0  \; ,
\ee
with solutions
\be \label{psi}
\psi(r) = \exp \Big(\pm A  \int \frac{dr}{f(r)}\Big) \; .
\ee
As a check, notice that $r$ is a function of the Lorentz invariant combination $X \cdot X = \psi^2(r)$, 
and so is 
\be \label{new metric}
G^{00} = -G^{11} = -\frac{A^2}{f} \psi^2 \cosh^2 + f \psi' {}^2
\sinh^2 = - A^2 \frac{\psi^2(r)}{f(r)} \; ,
\ee
as desired.

This is the main result of our paper:
we have shown that the two-dimensional, non-angular part of the 
black hole metric can be put into
a manifestly Lorentz invariant, and conformally flat, form.
Moreover, since the angular part of the metric depends only on $r$ which is
a function of $X \cdot X$, even that part is Lorentz invariant.

This discussion would be incomplete without showing that the 
metric in $(T,X)$ coordinates is in fact finite and regular at the
horizon.
The main ingredient is to assume a finite and non-zero surface gravity at the
horizon, which can be shown to imply 
\be \label{near horizon}
f(r) \simeq B (r - r_h)  \; , \qquad \mbox{for } r \simeq r_h \; ,
\ee
where $B$ is a non-vanishing constant. 
It can be further shown that choosing the plus sign in 
eq. \eqref{psi} and $A=B/2$ ensures that the new metric
is finite and regular at the horizon, allowing the $(T,X)$ coordinates
to continue beyond the horizon, thereby reproducing the 
Kruskal-like bifurcation structure.
These last steps of the argument are given in
\cite{RW1, RW2}. We summarize them in abbreviated form in
Appendix A.

\section{Concluding remarks}
For any static, spherically symmetric black hole with finite surface gravity, we have provided an explicit construction of Kruskal-like coordinates that realize the underlying time-translational invariance manifestly as a 2D Lorentz symmetry, and that make the metric manifestly regular at the horizon, thus allowing extrapolation past it. Although our results are implicitly already contained in \cite{RW1, RW2}, our emphasis on 2D Lorentz invariance allowed us to give a quicker, more direct derivation.

Schematically, our construction goes as follows:
\begin{itemize}

\item For any given initial metric in Schwarzschild-like coordinates (eq.~\eqref{metric}), one extracts the value of $B$ from the near-horizon behavior of $f(r)$ (eq.~\eqref{near horizon}).

\item The desired Kruskal-like coordinates are given by eqs.~\eqref{T}
  and \eqref{X}, with vanishing $\varphi(r)$, and with
\be
\psi(r) = \exp \Big(\frac B 2  \int \frac{dr}{f(r)}\Big) \; .
\ee

\item
The 2D metric in the new coordinates is manifestly regular at the horizon and at the origin, and takes the conformally flat, Lorentz invariant form
\be
G_{\alpha \beta} = \frac{4 f(r)}{B^2 \psi(r)} \, \eta_{\alpha \beta} \; .
\ee
The conformal factor is Lorentz invariant, because $r$ is. The Lorentz-invariance of $r$ further implies that the angular part of the metric,
\be
ds^2 \supset \rho^2(r ) \, d\Omega^2 \; ,
\ee
is also Lorentz-invariant. 
\end{itemize}

When one continues the metric past the horizon in these new coordinates, the emerging structure is the familiar one from the Kruskal extension of the Schwarzschild solution, with a bifurcation point, a light-cone, and four `wedges', with the right one corresponding to the region covered by the original Schwarzschild-like coordinates. The manifest $T$-inversion and $X$-inversion symmetries of the new metric, tell us that  the left and right wedges have the same geometry, and so do the upper and lower ones. In particular, since by construction the new coordinates cover all of the right wedge---our horizon is supposed to be the outermost zero of $f(r)$, and so there are no singularities in the integral in \eqref{psi} outside the horizon---they also cover all of the left wedge. Our new coordinate system can only break down in the upper and lower wedges, in a Lorentz invariant and $T$-inversion invariant fashion, that is, along the two branches of an hyperbola. In the Schwarzschild black hole case, it does so at the physical singularity, corresponding to the hyperbola \eqref{hyperbola}.

\section*{Acknowledgements}
We would like to thank Matt Kleban, Federico Piazza and Bob Wald for useful discussions. This work is supported by NASA under contract NNX10AH14G, and by the DOE under contracts 
DE-FG02-11ER1141743 and DE-FG02-92-ER40699.


\appendix

\section{Regularity of the new metric at the horizon}

In the construction of the $(T,X)$ coordinates in
Sec. \ref{2DLorentz}, we 
made no assumptions about the function $f(r)$ that characterizes our
original metric. 
Let us assume here that it features a finite surface gravity at the horizon. The reason is the following. In order for  there to be an origin $(T,X) =0 $ serving the role of  the center of our Lorentz transformations (item 4), we need the Killing vector to satisfy (see e.g.~\cite{Weinberg})
\be
\Xi^{\alpha} (0) = 0 \; , \qquad \nabla_{[ \alpha} \Xi_{\beta ] } (0) \neq 0 \; .
\ee
These conditions are covariant and can be analyzed in the original, Schwarzschild-like coordinate system. In particular the second one reads
\be
\nabla_{[\mu} \xi_{\nu]} \nabla^{[\mu} \xi^{\nu]}   \to \mbox{\rm finite}, \qquad \mbox{for } r \to r_h \; .
\ee
But this combination is precisely, up to factors of 2 and an overall square root, the definition of the horizon's surface gravity $\kappa$ 
\cite{Wald} \footnote{Notice that the black hole's horizon is a Killing horizon for $\xi^\mu$, since
\be
\xi^\mu \xi_\mu = -f(r) \to 0 \quad \mbox{for } r \to r_h \; .
\ee}.
The finiteness of $\kappa$ translates directly into a condition for $f$:
\begin{align}
\nabla_{[\mu} \xi_{\nu]} \nabla^{[\mu} \xi^{\nu]} & = g^{\mu \rho} g^{\nu \sigma} \di_{[ \mu} \xi_{\nu ]} \di_{[\rho} \xi_{\sigma ] } \\
& \propto f' {}^2 \to \mbox{\rm finite}, \qquad \mbox{for } r \to r_h \; .
\end{align}

We can now use this assumption to impose manifest regularity of the metric at the horizon (item 3).
A finite $f'$ at the horizon means that, close to the horizon, $f(r)$
can be approximated as
$f(r) \simeq B (r - r_h)$, for $r \simeq r_h$ (eq. \eqref{near
  horizon}). 
This is the standard behavior for horizons that are locally equivalent to Rindler's
\footnote{Extremal horizons are a notorious exception to this near-universal behavior.}. 
If we want the new metric components \eqref{new metric} to have  a finite (non-infinite and non-vanishing) limit at the horizon, we need  $\psi$ to behave like $\sqrt {f(r)}$ close to the horizon. Plugging \eqref{near horizon} into \eqref{psi} we get
\be
\psi(r) \simeq C (r- r_h)^{\pm A /B} \; , \qquad \mbox{for } r \simeq r_h \; ,
\ee
where $C$ is an integration constant. This is the same as the near-horizon behavior of $\sqrt {f(r)}$ if we choose the plus sign and 
\be
A = \sfrac12 B \; .
\ee
These choices uniquely specify our new coordinate system up to a
common arbitrary normalization of $X^{\alpha}$, corresponding to the
integration constant $C$. 
With these choices, the new metric (eq. \eqref{new metric}) is
manifestly finite at the horizon.
As a check, notice that the Kruskal extension of the Schwarzschild solution obeys all of the above with $B=1/r_h$.

We leave checking the other desired properties to the reader. 
Notice that regularity of the metric is not just about the finiteness
of the metric coefficients when we approach the horizon. In
particular, if we want to continue our new coordinates past the
horizon, we want the metric in the new coordinates to have at least
finite (non-infinite) first derivatives at the horizon. In Appendix
B, we collect a number of properties of $f(r)$ and $\rho(r)$ that follow from the regularity of {\em curvature invariants} at the horizon, and which guarantee that our new metric is indeed regular enough to be extrapolated past the horizon. The bifurcation structure---the presence of a `white hole' horizon, intersecting with the `black hole' one---just follows from time-inversion invariance, which is clearly a symmetry of our spacetime (manifest in both coordinate systems), and from the regularity of the metric at the origin.

\section{Regularity of curvature invariants at the horizon}
Let us use units in which the horizon is at $r = 1$.
The horizon is a null surface with $\rho = {\rm const} \neq 0$, assuming a non-vanishing surface area. The normal to such a surface,
$\di_\mu \rho$, is thus null, that is, 
\be
f \rho' \,^2 \to 0 \qquad (r \to1) \; .
\ee

Consider the curvature invariant $R_{\mu\nu} R^{\mu\nu}$, which should
be regular (non-singular) at the horizon, and which for the metric \eqref{metric} reads
\begin{align}
R_{\mu\nu} R^{\mu\nu} = & \frac2{\rho^4}{\left[-\rho  \left(f' \rho '+f \rho ''\right)-f\rho '^2+1\right]^2} \\
   & +\frac{1}{4 \rho^2} \left[\rho f''+ {2 f' \rho '} \right]^2\\
   & +\frac1{4 \rho ^2} {\left[\rho f''+2 f' \rho '+4 f \rho ''\right]^2} \; .
\end{align}
It is a sum of squares, which means that all combinations in brackets
should be separately regular. Dropping the 1 and the $f\rho '^2$ in
the first bracket, since they are regular by themselves, we can take suitable linear superpositions
of such combinations to show that $f'', f' \rho'$, and $f\rho''$ should
all be regular at the horizon. 
If we further assume that the surface
gravity $\kappa = f'/2$ is finite {\it and} non-zero, we find
that $\rho'$ is regular at the horizon as well.
(The surface gravity $\kappa$ is defined by $\kappa^2 = -\sfrac12 \nabla_\mu \xi_\nu \, \nabla^\mu \xi^\nu$, 
where $\xi^\mu$ is the time-translational Killing vector, eq.~\eqref{Killing}. For the metric \eqref{metric}, we get $\kappa = \sfrac12 f'$.)



\end{document}